\documentclass[a4paper]{article}

\usepackage{icrc2013}
\usepackage{subfigure} 
\usepackage{amstext}
\usepackage{url}

\newcommand{\hermes}{\texttt{HERMES}}
\newcommand{\hermess}{\texttt{HERMES}~}
\newcommand{\etal}{\MakeLowercase{\textit{et al. }}} 

\usepackage{color}

\title{ HERMES: a Monte Carlo Code for the Propagation of Ultra-High Energy Nuclei}

\shorttitle{M. De Domenico \etal The HERMES code for the propagation of UHE nuclei}

\authors{
Manlio De Domenico$^{1,2,3}$,
Haris Lyberis$^{4}$,
Mariangela Settimo$^{5,6}$
}

\afiliations{
$^1$ Laboratorio sui Sistemi Complessi, Scuola Superiore di Catania, Via Valdisavoia 9, 95123 Catania, Italy \\
$^2$ Dipartimento di Fisica e Astronomia, Universit\'a degli Studi di Catania, and Istituto Nazionale di Fisica Nucleare, Sez. di Catania, Via S. Sofia 64, 95123 Catania, Italy\\
$^5$ Universit\"at Siegen, Germany \\
$^4$ Universidade Federal do Rio de Janeiro, Brazil \\
\scriptsize{
$^{3}$ now at:  Departament d'Enginyeria Inform\'atica i Matem\'atiques, Universitat Rovira i Virgili, Tarragona, Spain \\
$^{6}$ now at:  Laboratoire de Physique Nucl\'eaire et de Hautes Energies (LPNHE), Universit\'es Paris 6 et Paris 7, CNRS-IN2P3, Paris, France. \\
}
}

\email{manlio.dedomenico@ct.infn.it}

\abstract{
Although the recent experimental efforts to improve the observation of Ultra-High Energy Cosmic Rays (UHECRs) above $10^{18}$~eV, the origin and the composition of such particles is still unknown. In this work, we present the novel Monte Carlo code (\hermes) simulating the propagation of UHE nuclei, in the energy range between $10^{16}$ and $10^{22}$~eV, accounting for propagation in the intervening extragalactic and Galactic magnetic fields and nuclear interactions with relic photons of the extragalactic background radiation. 
In order to show the potential applications of \hermess for astroparticle studies, we estimate the expected flux of UHE nuclei in different astrophysical scenarios, the GZK horizons and we show the expected arrival direction distributions in the presence of turbulent extragalactic magnetic fields.

A stable version of \hermess will be released in the next future for public use together with libraries of already propagated nuclei to allow the community to perform mass composition and energy spectrum analysis with our simulator.
}


\begin{document}
\maketitle

\section{Introduction}

A final answer about the origin and the composition of ultra-high energy cosmic rays (UEHCR) is still missing. Several models have been proposed for the acceleration of UHECR (see \cite{nagano2000observations, bhattacharjee2000origin} and Ref. therein for a review) and it is generally accepted that the candidate sources are extragalactic and trace the distribution of luminous matter on large scales \cite{waxman1997signature}. Even the observed suppression of UHECR, due to their propagation in the Universe, is still debated: in fact, UHECR of extragalactic origin with energy above 100\,EeV (1\,EeV $= 10^{18}$ eV) could be subjected to a strong attenuation because of their interaction with relic photons of the extragalactic background radiation. Recently, the Pierre Auger Collaboration reported a suppression of the spectrum above 40\,EeV with significance greater than 20 standard deviations \cite{settimo2012spectrum,abraham2010measurement}, improving previous measurements \cite{abraham2008observation,abbasi2008first}. Such results are compatible with the existence of the GZK effect \cite{zatsepin1966upper}, although the alternative scenario spectrum cutoff directly at the source also can not be excluded. Realistic simulations of production and propagation mechanisms are required to shed light on the nature of UHECR, trying to avoid the limitations of the continuous energy loss approximation.


In this work, we present the novel Monte Carlo code (\hermes) simulating the propagation of UHE nuclei, in the energy range between $10^{16}$ and $10^{22}$~eV, accounting for all such factors \cite{dedomenico2011thesis,dedomenico2013hermes}. 
In order to show the potential applications of HERMES for astroparticle studies, we estimate the expected flux of UHE nuclei in different astrophysical scenarios, the GZK horizons and we show the expected arrival direction distributions in the presence of turbulent extragalactic magnetic fields.

\section{Simulating UHECR propagation with HERMES}\label{sect:physics}

The \hermess propagation code allows the modeling and the simulation of i) the cosmological framework, ii) the cosmic background radiation (microwave, infrared/optical and radio), iii) the regular component of the Galactic magnetic field and the irregular component of both the Galactic and the extragalactic magnetic fields, iv) the cross sections describing the interactions between UHE nuclei and photons of extragalactic background radiation, v) the production of secondary particles because of such interactions. In the following, we will briefly describe such a framework, to provide the reader with the necessary tools to understand the parameterizations and the energy-loss equation adopted in our Monte Carlo code.

Motivated by up-to-date observations, we have chosen a general Friedmann's Universe, defined by a Friedmann-Robertson-Walker metric, to be the cosmological framework in \hermes. Our simulator is able to propagate particles in a $\Lambda$CDM Universe, with several cosmological tunable parameters. A study of the impact of cosmology on propagation of UHECR using \hermess was performed in~\cite{dedomenico2012influence}.

Concerning the modeling of sources, in \hermess is possible to use an homogeneous distribution of sources or any custom non-uniform distribution. If we indicate with $\mathcal{L}$ source luminosity, the injection spectrum of UHECR is modeled by $Q(z,E)=\mathcal{H}(z)Q(0,E)$, where $Q(0,E)=\mathcal{L}\mathcal{N}E^{-\gamma}$, being $\gamma$ the injection index, $\mathcal{N}$ a normalization factor and $\mathcal{H}(z)$ the source evolution factor depending on redshift $z$. Evolutions according to star formation rate (SFR), gamma-ray burst (GRB), active galactic nuclei (AGN) and quasi-stellar object (QSO) are available, as well as evolution of the form $\mathcal{H}(z)=(1+z)^{m}$.

For the propagation of UHECR nuclei, in \hermess we adopt the blackbody model with temperature $T_{0}\simeq 2.725$~K for cosmic microwave background (CMB). The semi-analytical ``model D''  proposed by Finke et al \cite{finke2010modeling}, modeling the star formation rate introduced by Hopkins and Beacom \cite{hopkins2006normalization} is adopted for cosmic infrared/optical background (CIOB). The model proposed in \cite{protheroe1996new} is adopted for the universal radio background (URB) and its evolution. Uniform redshift evolution is assumed for CMB, while both ``base-line'' and ``fast'' models are available for CIOB \cite{stecker2006intergalactic}.

In \hermes, the propagation of charged particles through magnetic fields is based on the numerical integration of the equation of motions obtained in the ultra-relativistic approximation. In practice, charged particles accelerating in a magnetic field lose energy because of the emission of synchrotron radiation: in the case of light particles as electrons or positrons, such energy loss should be taken into account during the propagation, whereas for heavier particles as protons it is negligible.

While the trajectory of a charged particle along the regular field is deterministic, i.e. for a given initial condition only one solution to the equations of motion exists, the trajectory of a particle through the turbulent field is stochastic, thus not unique, and it depends on the features of the irregular field as its r.m.s. strength and its coherence length. Unfortunately, we have no exact knowledge of both galactic and extragalactic magnetic fields and, as a consequence, the investigation of charged particles propagation through our galaxy and intergalactic space, respectively, should be based either on empirical or theoretical models and numerical simulations. For the simulation of the irregular component of the magnetic field, we adopt in \hermess the approach proposed by Giacalone and Jokipii \cite{giacalone1999transport}, based on a local step-by-step simulation of the turbulent Kolmogorov-like field. For the simulation of the regular component of the Galactic magnetic field, several models are included, allowing also for bisymmetric (BSS) or axisymmetric (ASS) modes. 


Finally, let us describe how nuclear interactions with relic photons of the extragalactic background radiation are taken into account. During their propagation, photons, neutrinos and nuclei $(A,Z)$ (electric charge, mass) with injection energy $E_{i}$, generally undergo interactions with background photons. UHECR that reach the Earth are therefore detected with a degraded energy $E_{f}<E_{i}$, depending on the type of interactions they were subjected to and on the distance between the source and the Earth. In \hermes, we describe the energy loss of non-stochastic processes in a unit interval of $z$ in terms of equations like
\begin{eqnarray}
\frac{1}{E}\frac{dE}{dz}=-\beta(z,E)\frac{dt}{dz},
\end{eqnarray}
where
\begin{eqnarray}
-\frac{dt}{dz}&=&\frac{1}{H_{0}(1+z)}[ \Omega_{M}(1+z)^{3} \nonumber\\
&+& \Omega_{\Lambda} + (1-\Omega_{M}-\Omega_{\Lambda})(1+z)^{2} ]^{-\frac{1}{2}}
\end{eqnarray}
is the general metric element accounting for the cosmological expansion \cite{engel2001neutrinos, ave2005cosmogenic, stanev2009high}. The function $\beta(z,E)$ is related to the cooling rate of the UHE particle and it depends on the particular energy loss process considered. It is proportional to the inverse of the mean free path and depends on the density of background photons and their energy, on the energy of the UHECR and on the cross section of the interaction under investigation. In the case of nuclei, it also depends on the nuclear mass and charge. Thus, the total energy loss rate is obtained by
\begin{eqnarray}
\label{def-energylosseq}
\frac{1}{E}\frac{dE}{dz} = -\frac{dt}{dz}\sum_{process}\beta_{proc}(z,E),
\end{eqnarray}
where the sum is extended to all interactions acting during the propagation. In \hermes, we include only those interactions which have a significant impact on the propagation of UHECR: i) \emph{adiabatic loss}, due to the expansion of the universe; ii) \emph{positron/electron pair production}; iii) \emph{photo-pion production}, involving the creation of one or multiple pions; iv) \emph{photodisintegration}, involving the fragmentation of the original nucleus, with the creation of lighter  nuclides (generally referred to as \emph{fragments}). For the implementation of these interactions we refer to~\cite{dedomenico2013hermes} and the one regarding the development of the electromagnetic cascade to \cite{settimo2012eleca}.

The resulting interaction length, in the cases of both proton and iron, are in perfect agreement with recent literature, e.g.~\cite{harari2006ultrahigh}, with small differences related to the different CIOB adopted.

We conclude this section with the production of secondary particles, namely photons and neutrinos, allowing multi-messenger analysis. We have discussed the production of electron/positron pairs and of secondary pions. Produced UHE photons and pairs interact with the extragalactic background photons, participating to the electromagnetic cascade generated by the primary nucleus. Conversely, in the case of photo-meson production, pions have small lifetime, of the order of $10^{-16}$~s for $\pi^{0}$ and $10^{-8}$~s for $\pi^{\pm}$: thus, we neglect their propagation because they quickly decay to new secondary particles, which can decay to other particles (as in the case of secondary muons) generating a cascade of electrons, positrons, photons and neutrinos. In \hermes, we consider all the main decay channels involving the production of a single pion and we include the $\beta-$decay of neutrons.

The propagation of UHE photons produced by neutral pions, and the consequent pairs, are performed with \texttt{EleCa}~ and its description is beyond the scope of the present work. We refer to \cite{settimo2012eleca,settimo2013eleca} for further details. Neutrinos, produced by the decay of charged pions and $\beta-$decay of neutrons undergo interactions only through the weak nuclear force. Thus, propagation and energy loss of neutrinos, can be easily described by energy loss equation (\ref{def-energylosseq}), considering only the adiabatic energy loss rate. Because of such features, neutrinos are likely to traverse the extragalactic space, even for cosmic distances, without interacting with background photons or interstellar medium, and without being deflected by magnetic fields: characteristics that makes neutrinos the ideal candidates for particle astronomy.

\section{Applications at ultra-high energies}\label{sect:applications}

In this section we briefly discuss some applications to show the potentiality of \hermess for studying UHECR, including the comparison between results obtained with \hermess and those either from other propagation codes available in the UHECR community or from observation.

\begin{figure}[!t]
	\centering
	  \includegraphics[width=8.5cm]{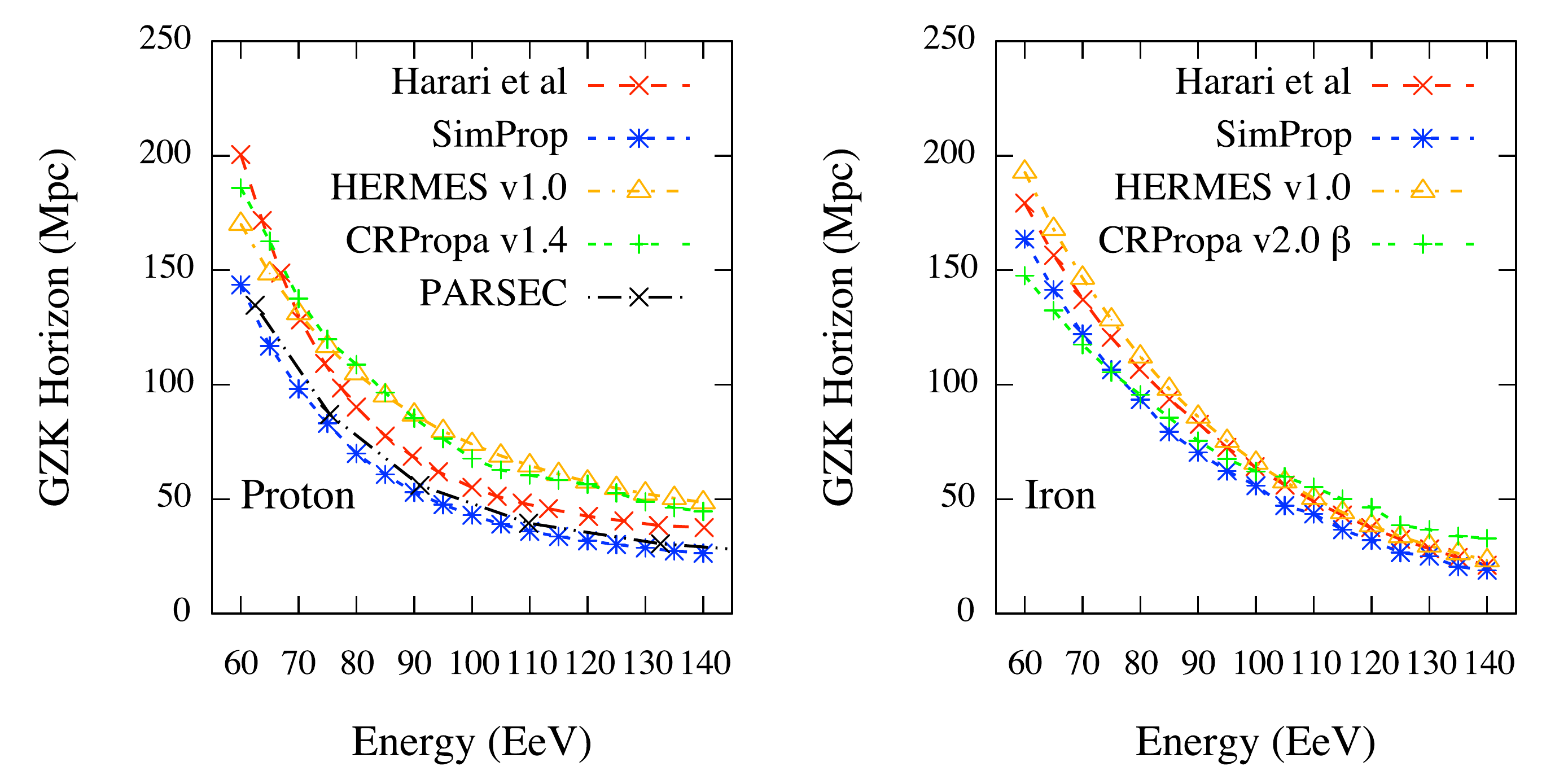}
	\caption{GZK horizon estimated in the case of protons (left panel) and iron nuclei (right panel) injected with spectrum $E^{-2.7}$, as a function of the energy threshold at Earth. Results from CRPropa~v2.0$\beta$, the up-to-date version of the Monte Carlo code simulating the 3D propagation of nuclei in a magnetized Universe~\cite{sigl2011icrc,kampert2013crpropa}, and Harari \emph{et al} \cite{harari2006ultrahigh} are shown for reference.}
\label{fig-hermes-comp4}
\end{figure}

\begin{figure}[!t]
	\centering
   \includegraphics[width=8.5cm]{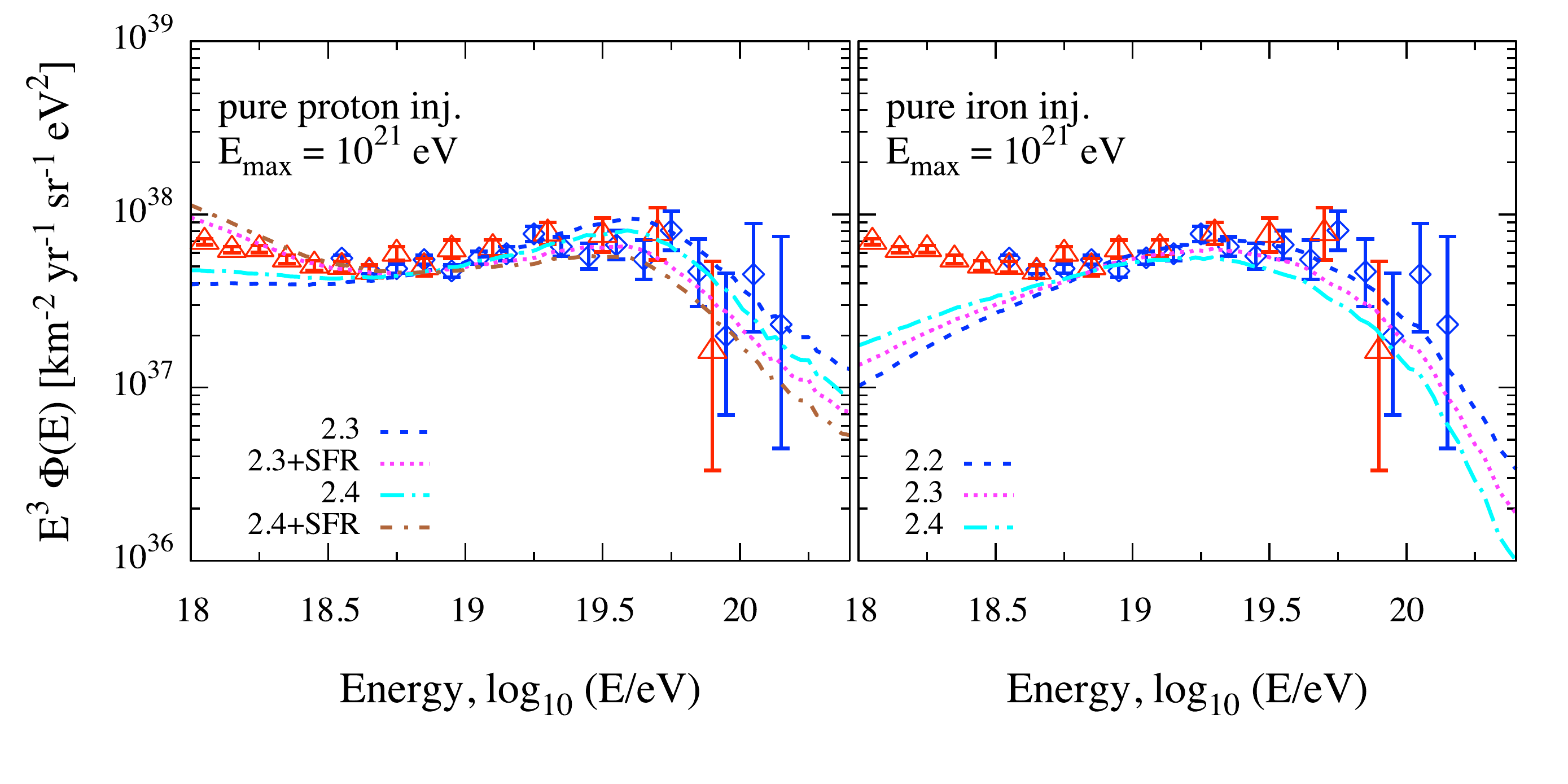}
  \caption{Expected all-particles energy spectra obtained from \hermess for different astrophysical scenarios, compared to observations reported by HiRes Collaboration (see the text). The legend indicates the spectral index at the source and the source evolution adopted (only star formation rate, in this case).}
   \label{fig:HERMES-spectra}
\end{figure}

The GZK horizon as a function of the energy threshold at Earth is a well-known results in literature~\cite{harari2006ultrahigh} and CRPropa\footnote{The version used here is dated September 2011.} allowing for comparisons. In Fig.\,\ref{fig-hermes-comp4} we show the GZK horizon of protons (left panel) and iron nuclei (right panel): the horizons obtained by \hermess are in agreement with those of CRPropa over the whole energy range under consideration, although for iron nuclei some differences are present at the lowest energy. 

Moreover, we estimate the expected energy spectra of UHECR at Earth in different astrophysical scenarios, involving evolution of sources, different spectral indices and mass composition at the source. The result, shown in Fig.\,\ref{fig:HERMES-spectra}, are compared against recent observations reported by the HiRes Collaboration \cite{abbasi2008first}. For sake of simplicity, we show only some representative spectra: a study of their goodness in reproducing the observed UHECR spectrum is beyond the scope of the present paper and it will be the subject of a future study.

\begin{figure*}[!t]
\centering
\subfigure[2MRS Catalogue]
   {
	  \includegraphics[width=6.6cm]{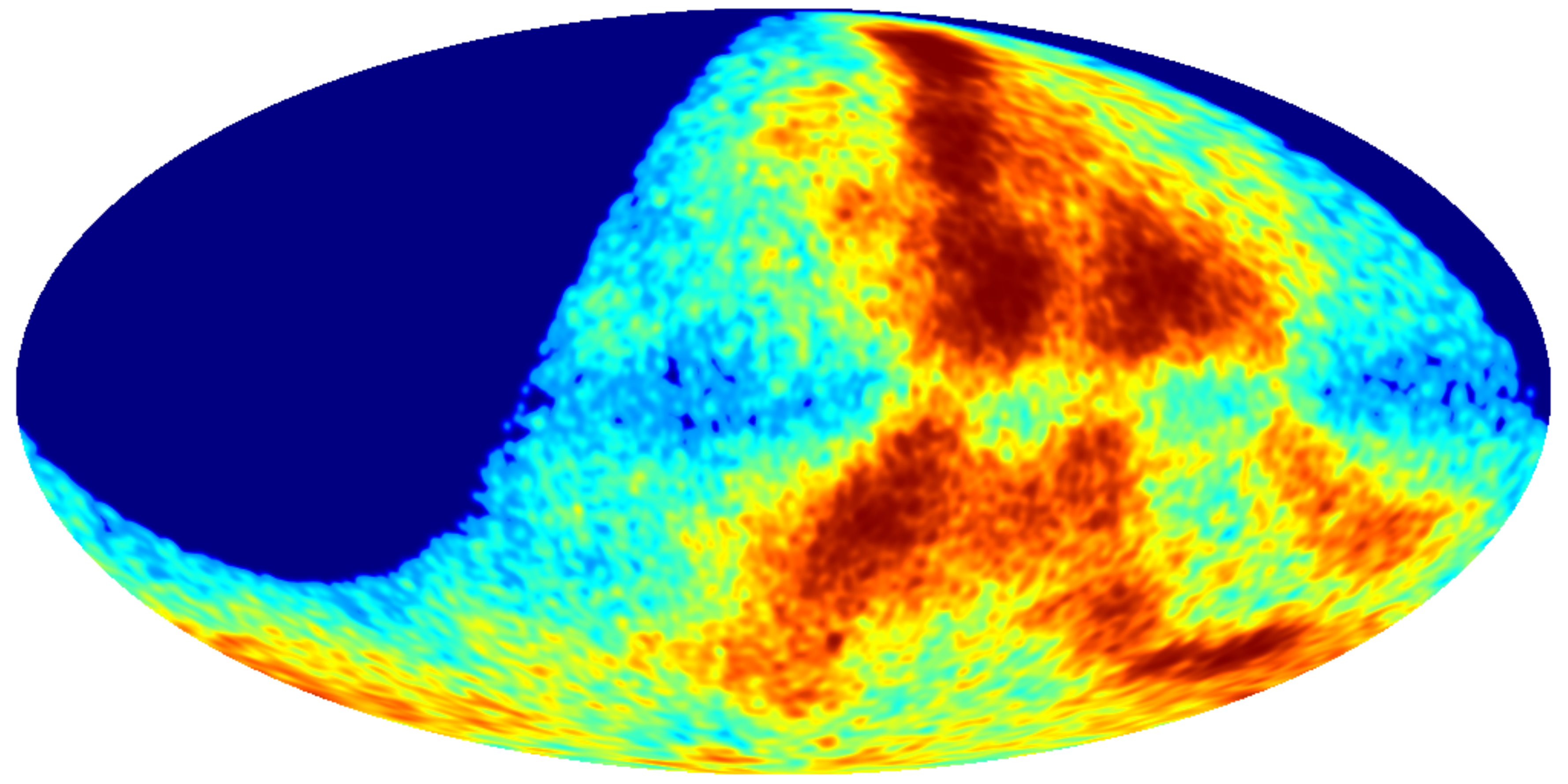}
	  \label{fig:skymap2MRS}
   }
   \hspace{5mm}
\subfigure[SWIFT58 Catalogue]
   {
	  \includegraphics[width=6.6cm]{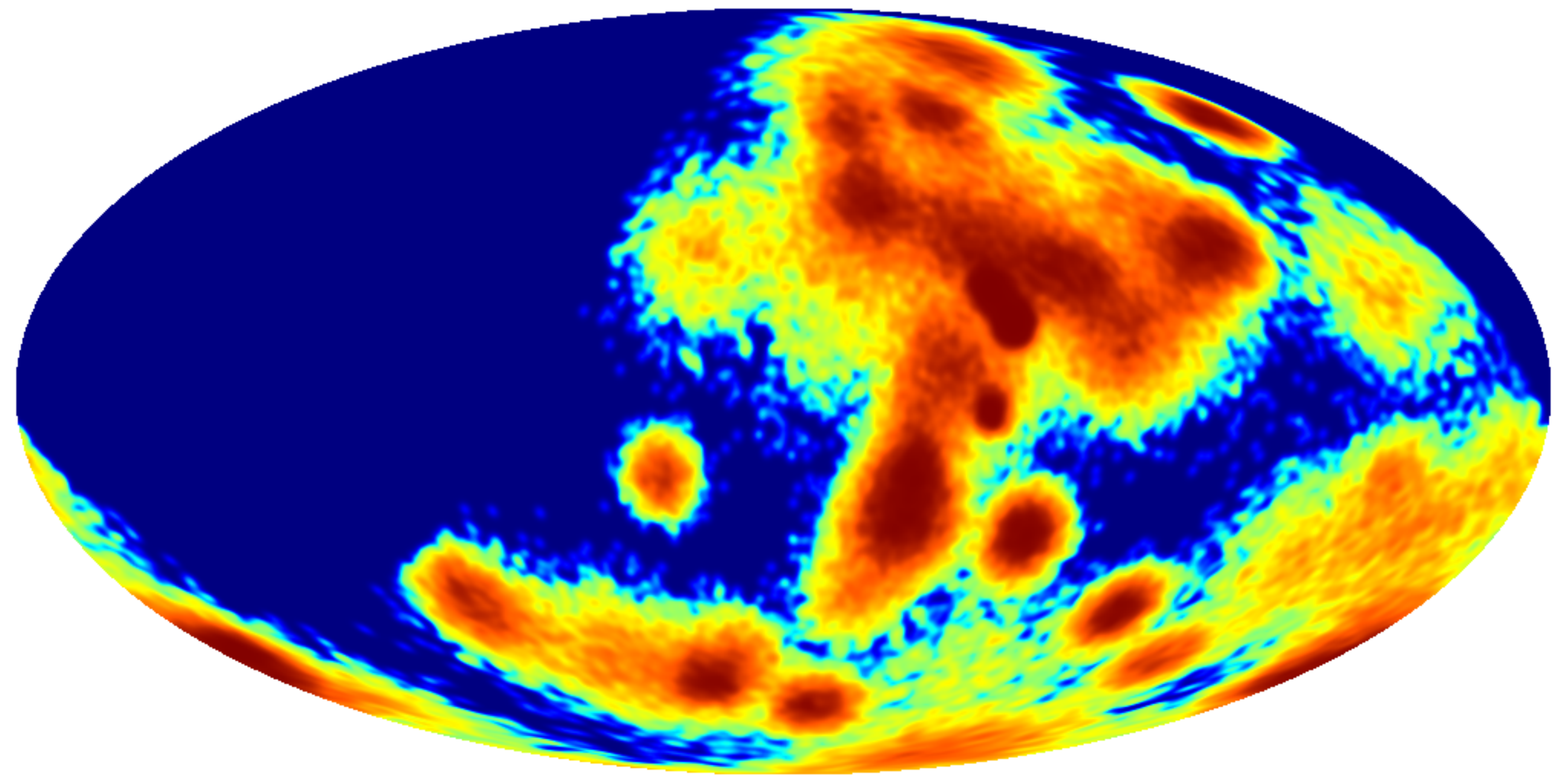}
	  \label{fig:skymapSWIFT}
   }
\subfigure[2MRS Catalogue + Isotropic]
   {
	  \includegraphics[width=6.6cm]{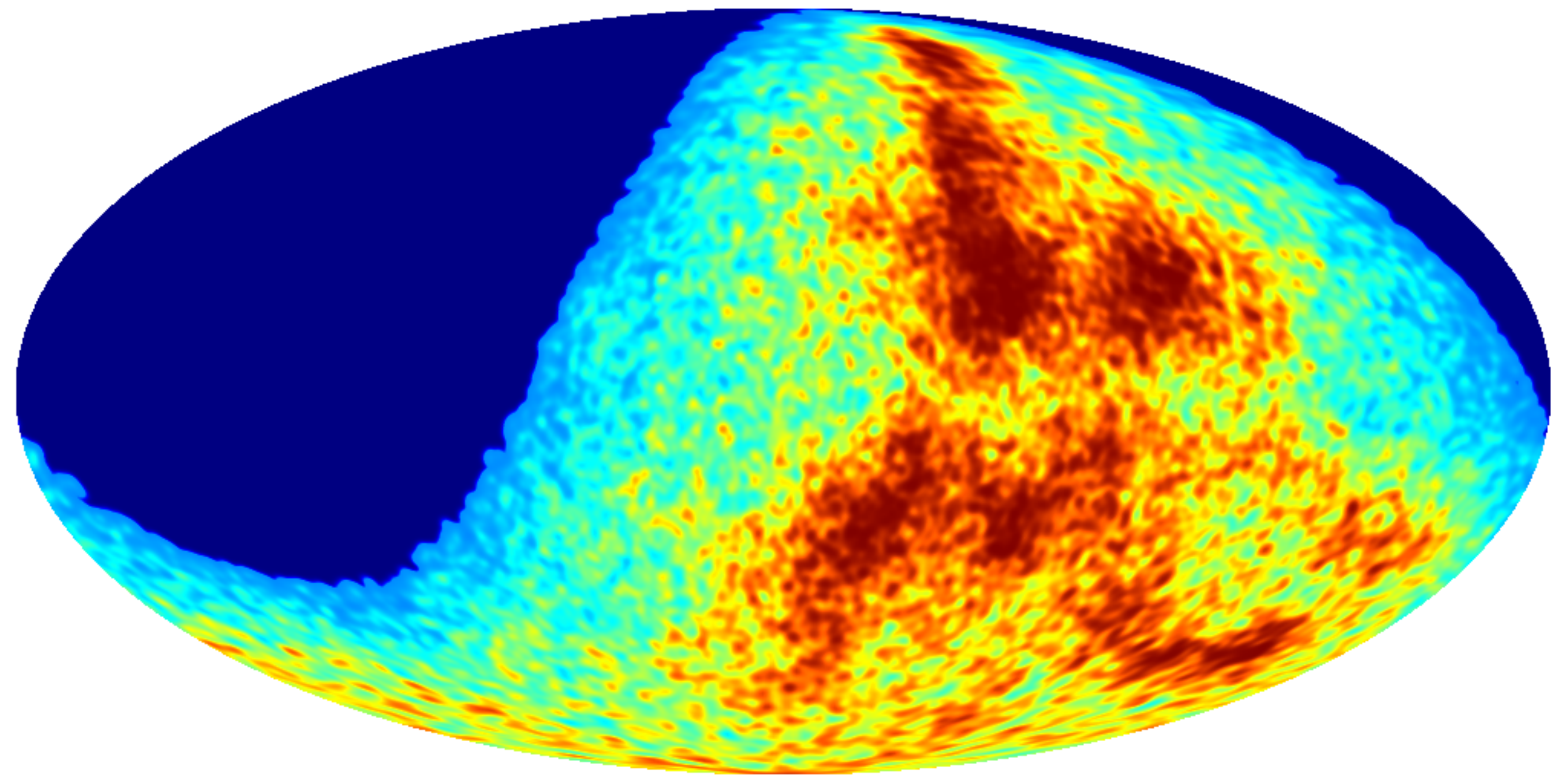}
	  \label{fig:skymap2MRSiso}
   }
   \hspace{5mm}
\subfigure[SWIFT58 Catalogue + Isotropic]
   {
	  \includegraphics[width=6.6cm]{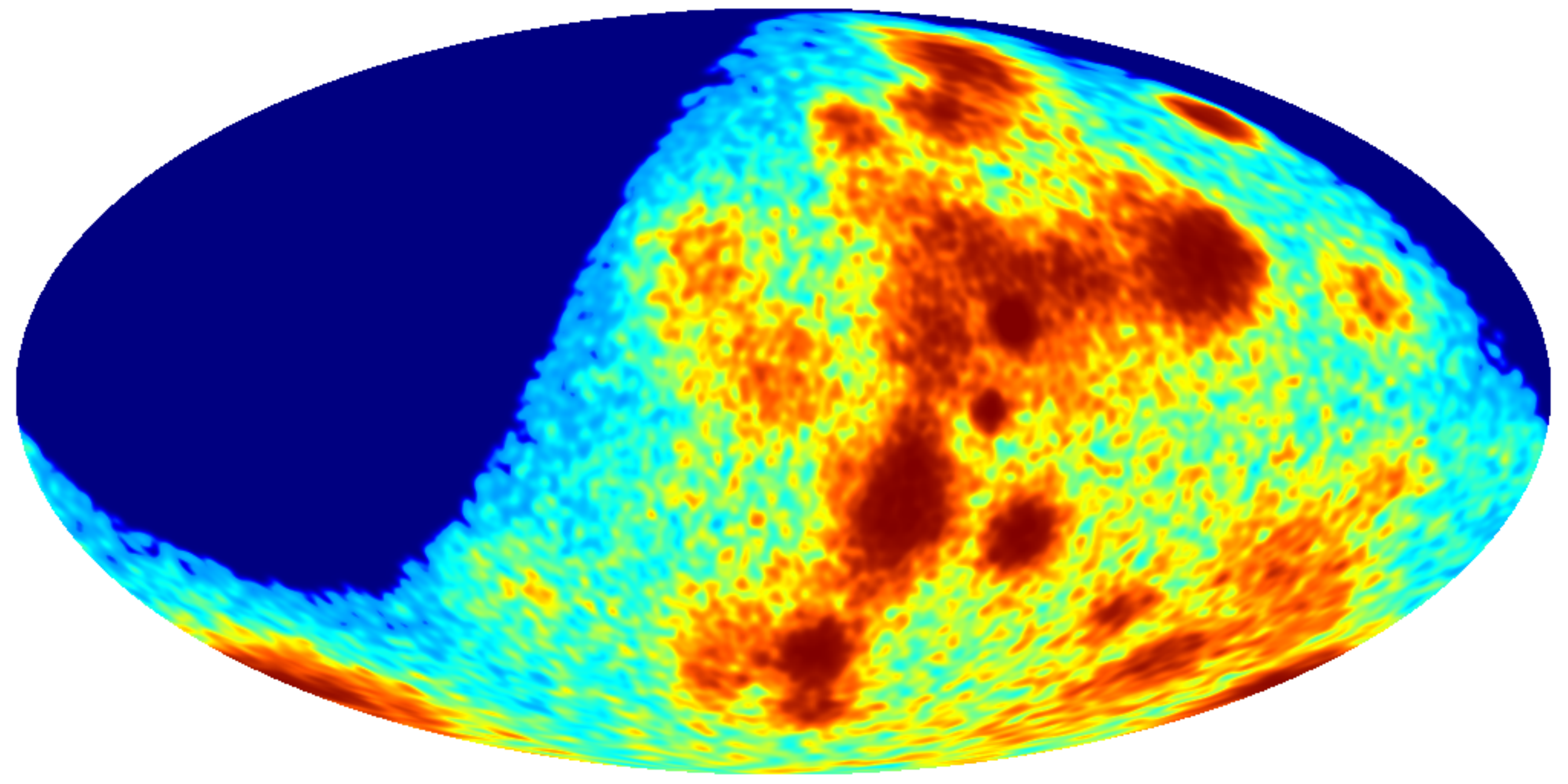}
	  \label{fig:skymapSWIFTiso}
   }

\caption{Skymaps, accounting for the Pierre Auger Observatory non-uniform exposure, of simulated UHE protons produced by nearby sources (within 200~Mpc) experiencing deflections due to an intervening extragalactic magnetic field. Galactic coordinates are shown. 2MRS (\ref{fig:skymap2MRS} and \ref{fig:skymap2MRSiso}) and SWIFT-BAT 58-months (\ref{fig:skymapSWIFT} and \ref{fig:skymapSWIFTiso}) are considered. See the text for further details.}
\end{figure*}

As a final application, we simulated protons from real candidate sources in the nearby Universe, with distance between 4 and 200~Mpc. We included the effect of deflections due to an intervening Kolmogorov-like extragalactic magnetic field with r.m.s. strength of 2~nG and coherence length of 1~Mpc. Moreover, we consider the case of absence of isotropic contamination and the case where simulation are contaminated with 56\% isotropic events, according to recent measurements of the Pierre Auger Collaboration \cite{auger2010correlation}. The resulting skymaps of simulated events, as they would be observed by accounting for the non-uniform exposure of the Pierre Auger Observatory, are shown in Fig.\,\ref{fig:skymap2MRS} and Fig.\,\ref{fig:skymap2MRSiso}, for candidate sources of UHECR from 2MASS Redshift Survey \cite{huchra20112mass} with magnitude ranging from -27.5 to -9.8, and in Fig.\,\ref{fig:skymapSWIFT} and Fig.\,\ref{fig:skymapSWIFTiso} for active galactic nuclei from SWIFT-BAT 58months \cite{swift2010}.

\section{Conclusion}


Realistic simulations of the propagation of UHECR might help to shed light on their origin and their nature. In this work, we presented \hermes, the \emph{ad hoc} Monte Carlo code we have developed to propagate UHECR in a magnetized Universe. We have briefly discussed the theoretical framework behind \hermes, involving the modeling of cosmology, magnetic fields, nuclear interactions between UHECR and relic photons of the extragalactic background radiation, and the production of secondary particles. The distribution of sources, their intrinsic luminosity, injection spectrum and evolution are tunable parameters in \hermes, allowing to simulate a wide variety of astrophysical scenarios and to investigate the impact of propagation on physical observable as the flux, or the chemical composition observed at Earth. 

We showed some representative applications validating the suitability of \hermess for astroparticle studies at the highest energies. More specifically, we estimated the surviving probability of UHE protons, the GZK horizons of nuclei, the all-particle spectrum observed at Earth in different astrophysical scenarios and the expected arrival direction distribution of UHECR produced from different catalogues of nearby candidate sources.

The major advantage in using \hermess is in its modularity, allowing high customization of involved physical and astrophysical parameters. In fact, it is possible, for instance, to add new models of extragalactic background radiations or nuclear interactions, according to up-to-date measurements.

Although a deeper analysis of correlation and intrinsic clustering is out of the scope of this paper, the results show how \hermess can be used for such purposes. Moreover, it is possible to investigate the compatibility between simulated scenarios and observation by coupling \hermess with other methods. For instance, it is possible to quantify the clustering signal in the arrival direction distribution with multiscale autocorrelation \cite{dedomenico2011multiscale} or to perform multi-messenger analysis including photons propagated with \texttt{EleCa}~ \cite{settimo2012eleca}. Another interesting application is to use the parameterization based on the generalized Gumbel distribution \cite{dedomenico2013reinterpreting} to perform detailed mass composition studies, as comparing the expected first and second momenta of the $\text{X}_{\text{max}}$ distribution from different scenarios against observations.

In the near future, we will release a stable version of our simulator for public use and, in the meanwhile, we will make available for the community libraries of propagated nuclei useful for mass composition and energy spectrum analysis.

\vspace*{0.5cm}

\footnotesize{{\bf Acknowledgment:\\}{
M.S. acknowledges support by the BMBF Verbundforschung Astroteilchenphysik and by the Helmholtz Alliance for Astroparticle Physics (HAP). H.L. acknowledges support by CAPES, grand n. CAPES-PNPD 2940/2011.
}}


\end{document}